\documentclass{article}

\usepackage{arxiv}

\usepackage[utf8]{inputenc} 
\usepackage[T1]{fontenc}    
\usepackage{hyperref}       
\usepackage{url}            
\usepackage{booktabs}       
\usepackage{amsfonts}       
\usepackage{nicefrac}       
\usepackage{microtype}      
\usepackage{lipsum}
\usepackage{indentfirst}
\usepackage{setspace}

\usepackage{graphics}      
\usepackage{graphicx}      
\usepackage{longtable}     
\usepackage{float}

\newcommand{\mat}[1]{\mbox{\boldmath{$#1$}}}
\newcommand{\beq}{\begin{equation}}
\newcommand{\eeq}{\end{equation}}
\newcommand{\bea}{\begin{eqnarray}}
\newcommand{\eea}{\end{eqnarray}}

\title{CONTRIBUTIONS OF A MODIFIED ELECTRODYNAMICS TO THE MOLECULAR BIOCHIRALITY}

\author{{\Large
  A. C. L. Santos,\thanks{alana.santos@aluno.uece.br}\,\,\,\, C. R. Muniz,\thanks{celio.muniz@uece.br}\,\,\,\, L. T. Oliveira\thanks{leonardo.tavares@uece.br}\,\,\, and \,\,J. T. Souza}\thanks{jefferson.thiago@uece.br}\\
{\Large Universidade Estadual do Cear\'{a}}\\
{\Large Faculdade de Educa\c{c}\~{a}o, Ci\^{e}ncias e Letras de Iguatu}\\
{\Large Av. Dario Rabelo, s/n,
CEP: 63.500-000, Iguatu, CE, Brazil}\\
}

\begin{document}
\maketitle
\vspace{1.0cm}
\doublespacing 
\begin{abstract}
{\large 
In this paper, we investigate the physical basis behind the molecular biochirality from the computation of a parity violation energy difference (PVED) in enantiomers of organic molecules (e.g., amino acids, which occur as levogyrous-type in nature), by considering the influence of fundamental interactions beyond the standard model of elementary particles and interactions.
Particularly, we study the role of a 4-D Chern‑Simons theory at the origin of this PVED, the Carroll‑Field‑Jackiw electrodynamics, which violates both Lorentz and parity symmetries. Then, we consider terrestrial and Jovian scenarios where the influence of a modified (effective) magnetic field generated by the planets on the molecules is taken into account in the calculation of PVED. Besides this quantity, we also calculate the relative quantity excess of
an enantiomer over the other in a thermal bath. Finally, we compare the obtained results with those ones from other models based on fundamental interactions.}

\vspace{0.5cm}
\noindent{{\large Keywords: beyond standard model, biomolecular chirality, Carroll‑Field‑Jackiw electrodynamics}}
\end{abstract}

\renewcommand{\thesection}      {\Roman{section}}
\maketitle
\newpage
\section{INTRODUCTION}
{\large
In the middle of the 19th century, Louis Pasteur discovered
a singular peculiarity of the nature. According to his
observations, it at times produces dextrogyrous (which
rotates the plane of polarization of light to the right) or
levogyrous (which rotates the plane of polarization of
light to the left) molecules, when they are synthesized in
the biological metabolism: the so-called biochirality. In
some fundamental molecules for the development of life,
the biochirality is lead to the extreme (via autocatalysis
mechanism), becoming homochirality. For example,
there are practically only L-amino acids in proteins and
D-sugars in DNA and RNA \cite{Gal}. In this sense, molecular chirality
plays an important role in the function performed
in biological processes so that enantiomeric compounds
may have antagonistic functions or secondary actions of one of their stereoisomers. Such differences are consequences
of the specificity of chiral recognition in many
biological processes, performed between chiral receptors
or reagents that selectively interact with one of the chiral
molecule enantiomers \cite{Gujarro}. The presence of enzymes and stereospecific
receptors in organisms leads to the most
diverse biological responses \cite{Gujarro, Caglioti}.

The origin of biochirality seems to be closely linked to
the origin of life, and some hypotheses have been proposed
to explain such a feature \cite{Gujarro, Caglioti, Pavlov}. For instance, the discovery
of the disproportion between enantiomeric
quantities, dextrogyrous, and levogyrous of organic molecules
in the Murchison meteorite introduced the possibility
that biochirality was induced in space and sown on
Earth, reinforcing the hypothesis of the panspermia \cite{Kvenvolden, Cronin, Glavin}.

The physical mechanisms underlying the emergence
of the biochirality in certain organic molecules are still not clearly elucidated \cite{Takahashi}. Basically, it remains to understand
which fundamental mechanisms are really
involved and how they make one enantiomer energetically
more favorable than the other one (and more stable,
therefore), causing the imbalance in their quantities, at
least in Earth. Thus, for decades, it has been discussed in
theoretical proposals that take into account contributions
of fundamental forces producing a parity violating energy
difference (PVED) between the enantiomers. In weak
interaction, which has proven to have no parity symmetry,
the PVED comes from the electron-nucleon interaction
through the helicity operator \cite{Dortaurra}. Or from electron neutrino
interaction, through weak neutral currents, in
which the neutrinos resealed from a supernova induces a
much higher PVED \cite{Bargueno1, Bargueno2}. And finally, in the electron-
WIMPS interaction, where it is significantly small \cite{Bargueno3}.

Regarding the gravitational interaction, based on the
Leitner and Okubo's gravitational potentials, in 1976,
Hari Dass defined a correction to the Newtonian potential
that allowed us to calculate a PVED associated to the
particle when it is attracted by a macroscopic mass \cite{Leitner, Dass}.
In turn, the extent of general relativity, which uses the
Chern‑Simons theory of electromagnetism, proposed in
2009, also presents parity violation \cite{Alexander}. Finally, it comes in
loop quantum gravity (LQG), one of the models that
gathers general relativity and quantum mechanics in
small scales, near the Planck scale \cite{Freidel}. However, it is
important to say that these energy differences have not
yet been proven.

The extended standard model is a theoretical scenario
in which investigations into violations of Lorentz and
CPT symmetries are carried out. The identification of
Lorentz's invariance violation in the usual standard
model represents a sign of the existence of a more fundamental
theory, valid at the scale of very high energies.
Thus, another mechanism that could be behind the enantiomeric
disproportion in organic molecules, still not considered
in the literature up to here, is that one due to the
electrodynamics sector of the extended standard model,
as the Carroll‑Field‑Jackiw (CFJ) one \cite{CFJ}. This 4-D
Chern‑Simons theory leads to both the Lorentz and parity
violations in the electrodynamics since it adds to the
usual Faraday‑Maxwell theory a sector with a static
background four-vector field permeating the whole
spacetime, whose coupling with the electromagnetic
fields is very tiny.

The parameters of the theory considered here, basically
the components of the four-vector background field,
are bounded by several experimental measurements,
involving both terrestrial and astrophysical systems, and
the diverse bounds found were recently synthesized in a
previous study \cite{Kostelecky}. It is worth to mention that the Earth-based
experimental bounds are much less restrictive than the ones obtained in astrophysics but are less subject to
the influence of experimental errors common to these latter.
We also remark that such violations in the Lorentz
symmetry are actually considered in several theories that
seek to describe the physics near the Planck's scale \cite{Kostelecky2, Horava, Bluhm}.

The present work seeks to find a relationship between
the natural surplus of an enantiomeric form over the
other in organic molecules and a modified electrodynamic
theory. The reasoning follows the studies that consider
physical mechanisms beneath that disproportion,
currently explained by theories that violate the parity
symmetry, as, for instance, the weak nuclear interaction.
The idea here is that the potential energy coming from
effective magnetic fields described by the CFJ theory,
interacting with the magnetic dipole moment of the
amino-acid molecule, is different for each enantiomeric
form, inducing therefore a PVED. This energy difference,
when one considers a thermal bath, leads to the mentioned
numerical disproportion. At last, we compare the
obtained results with those ones exhibited in literature.

The paper is structured as follows: In Section 2, we
will present the CFJ electrodynamics and compute the
PVED. In Section 3, we will apply these calculations to
the terrestrial and Jovian scenarios. Finally, in Section 4,
we will present the conclusions and close the paper.

\section{PVED IN THE CARROLL‑FIELD‑JACKIW ELECTRODYNAMICS}

The modified Maxwell's equations can be obtained from
the covariant Lagrangian density
\begin{equation}
{\displaystyle\mathcal{L}=-\frac{1}{4}F^{\mu\nu}F_{\mu\nu}+j_{\mu}A^{\mu}}+\frac{1}{2}{\epsilon}_{\mu\nu\rho\sigma}K^{\mu}A^{\nu}F^{\rho\sigma},\label{e1}
\end{equation}
where $A^{\mu}=(\phi, A)$ is the electromagnetic potential fourvector, $F^{\mu\nu}={\partial}^{\mu}A^{\nu}-{\partial}^{\nu}A^{\mu}$ is the antisymmetric electromagnetic strength tensor, and $j^{\mu}=(\rho, \mathbf{J})$ is the four-current.
The last term of Equation (\ref{e1}) contains $K^{\mu}$, a
postulated background uniform four-vector that pervades
the whole spacetime, modifying the usual Maxwell
Lagrangian to an odd 4-D Chern‑Simons electromagnetic
theory, the so called Carroll‑Field‑Jackiw electrodynamics \cite{CFJ}. This latter is a beyond standard model theory that
is gauge but not Lorentz and parity invariant due to the
presence of the mentioned background four-vector. It is
worth noticing that the corresponding theory in 3-D (planar)
spacetime respects these symmetries since the
Levi‑Civita tensor has three subindexes, and it contracts
only with $A^\nu F^{\rho\sigma}$, without the necessity of introducing a background field, therefore.

The corresponding equation of motion results in 
$(c = 1)$:
\begin{eqnarray}
    \nabla\cdot\textbf{E}&=&4\pi\rho-\textbf{K}\cdot\textbf{B},\\
    \nabla\cdot\textbf{B}&=&0,\\
    \nabla\times\textbf{E}&=&\displaystyle -\frac{\partial \textbf{B}}{\partial t},\\
    \nabla\times\textbf{B}&=&4\displaystyle\pi\textbf{J}-K_0\textbf{B}+\textbf{K}\times\textbf{E}+\frac{\partial\textbf{E}}{\partial t}\label{e2}.
\end{eqnarray}
We can notice that the only equations modified with
respect to the usual electrodynamics are those ones with
sources (Faraday and Ampère‑Maxwell laws).

Adopting a reference frame in which the background
four-vector $\displaystyle K^\mu = (K_0, \textbf{K})$ has only the temporal component $K_0$ (rest frame), we can write Equation (\ref{e2}) as: 
\begin{eqnarray}
    \nabla\times\textbf{B}=4\displaystyle\pi\textbf{J}-K_0\textbf{B}+\frac{\partial\textbf{E}}{\partial t},
\end{eqnarray}
and by considering $\textbf{B} = \nabla\times \textbf{A}$, we get:
\begin{eqnarray}
   \displaystyle \nabla\times(\textbf{B}+K_0\textbf{A})&=&4\pi\textbf{J}+\frac{\partial\textbf{E}}{\partial t},\\
   \displaystyle \nabla\times\textbf{\~B}&=&4\pi\textbf{J}+\frac{\partial\textbf{E}}{\partial t},\;\;\;\textrm{with}\;\;\; \textbf{\~B}=\textbf{B}+K_0\textbf{A}.
   \end{eqnarray}
The Maxwell's equations can now be written in terms of $\textbf{\~B}$ by choosing the Coulomb gauge, $\nabla\cdot{\bf A}=0$, as well as
static electromagnetic fields, in the vacuum. On the other
hand, there is a difference in the expected value of the
Hamiltonian associated with the interaction between the
microscopic magnetic dipole moment $\mat\mu_a$ of the molecules
L and R and the redefined macroscopic magnetic field,
given by:
\begin{eqnarray}
  \displaystyle \langle H_R\rangle=-\langle\Psi|\displaystyle{\mat\mu}_a\cdot\textbf{\~B}|\Psi\rangle=-\langle\Psi|{\mat\mu}_a\cdot\textbf{B}+K_0{\mat \mu}_a\cdot\textbf{A}|\Psi\rangle.
\end{eqnarray}
By realizing a parity operation, one gets:
\begin{eqnarray}
   \displaystyle \langle H_L\rangle=-\langle\Psi|{\mat{\mu}}_{a}\cdot\textbf{B}- K_0{\mat{\mu}}_{a} \cdot\textbf{A}|\Psi\rangle,
\end{eqnarray}
The sign change after the parity operation is due to the
axial nature of both the magnetic field and dipole
moment vectors, as well as of the polar nature of the electromagnetic
potential three-vector. The PVED will be,
therefore
\begin{eqnarray}
  \displaystyle \Delta U=\langle H_L\rangle-\langle H_R\rangle=2\langle\Psi|K_0{\mat\mu}_a\cdot\textbf{A}|\Psi\rangle.
\end{eqnarray}
Assuming a spherical macroscopic body with magnetic dipole moment given by ${\bf m}$, we have that the expected value of the potential vector $\langle{\bf A(r)}\rangle$ will be
\begin{eqnarray}
  \langle \textbf{A}(r)\rangle=\frac{\mu_0}{4\pi}\textbf{m}\times\frac{\textbf{r}}{r^3},
\end{eqnarray}
so that
\begin{eqnarray}
   \Delta U =\displaystyle\frac{\mu_0}{2\pi}\mu_a K_0\frac{|\textbf{m}\times\textbf{r}|}{r^3}\cos{\theta},\label{e3}
\end{eqnarray}
where $\theta$ is the angle between ${{\bf \mat\mu}}_{a}$ and $\textbf{A}$.

\section{PVED IN TERRESTRIAL AND JOVIAN SCENARIOS}

From equation (\ref{e3}), an upper limit for PVED will be:
\begin{eqnarray}
 \displaystyle  \Delta U=\frac{\mu_0}{2\pi}\mu_a K_0 \frac{m}{r^2}.
\end{eqnarray}
For a terrestrial scenario, taking the upper limit for  as
being $K_0 = 5.0\times 10^{-11}$ m$^{-1}$, which is based on geomagnetic
arguments provided in Carroll et al \cite{CFJ}. and is more restrictive
than the terrestrial experiment-based bounds given
in Kostelecky and Russell \cite{Kostelecky}, $m = 8 \times 10^{22}$A.m$^2$, and
$r = 6,2 \times 10^6$m. Considering the amino-acid magnetic
dipole moment as being constituted of one unit of Bohr
magneton \cite{Wang},  $\displaystyle\mu_a \cong 1 \mathbf{\mu_B} = 9,2 \times 10^{-24} $JT$^{-1}$, we have that the maximum PVED will be
\begin{eqnarray}
   \Delta U \cong 1.2\times 10^{-12}\textrm{eV}.
\end{eqnarray}
This energy difference corresponds to an imbalance
between the enantiomeric quantities L and R in a thermal
bath at temperature $T\cong 300$ K given by
\begin{eqnarray}
   \Delta N=N_L - \displaystyle N_R=N_0\left[\exp{\left(-\frac{\langle H_L\rangle}{K_BT}\right)}-\exp{\left(-\frac{\langle H_R\rangle}{K_BT}\right)}\right],
\end{eqnarray}
with $N_0=N_L+ N_R$, resulting in an excess of L molecules with respect to R of
\begin{eqnarray}
   \displaystyle\Delta N \cong N_0\left(\frac{\Delta U}{K_BT}\right)\cong 10^{14}\;\;\textrm{molecules/mol. }
\end{eqnarray}

\begin{table}[ht]
 \centering \caption{PVED based on some models involving fundamental interactions, with our results on bold characters.}
    \label{Tabela1}
    \centering
\begin{tabular}{c|c}
   \hline\\
    INTERACTION  & PVED (eV) \\\\ \hline\\\\
    Electron-Neutrino (Supernova) \cite{Bargueno2} & $10^{-5}$ \\\\ 
     \textbf{Chern-Simons (Carrol-Field-Jackiw Electrodynamics)} & $10^{-9}-10^{-12}$ \\\\
    Chern-Simons (Gravitational) \cite{Bargueno4} & $10^{-10}-10^{-14}$ \\ \\
       Electron-Nucleon \cite{Dortaurra, Bargueno5} & $10^{-14}-10^{-16}$ \\\\ 
    Cosmological Neutrinos-WIMPS (in the most favorable case) \cite{Dortaurra} & $10^{-21}$ \\\\ 
    Electron-Cosmological Neutrino \cite{Bargueno1} & $10^{-26}$\\\\\hline
\end{tabular}  
\end{table}

With respect to the astronomical scenarios, if one extends to them the Earth-based bound for the $K_0$ component, we can obtain the PVED of organic molecules that eventually have passed near some of these astronomical objects. For instance, by considering a Jovian scenario, where the gaseous giant is source of an powerful magnetic field
with $m\cong 10^{27}$Am$^2$, and radius of $r\cong 7 \times 10^7$m, we have that the PVED bound for amino-acids at its surface is of the order of $10^{-9}$eV.

\section{CONCLUSION}

We have made an unprecedented proposal of considering
an extended theory of the electrodynamics, the
Carroll‑Field‑Jackiw theory, in order to explain the origin of the biochirality in organic molecules such as
amino acids, observed in Earth as well as in some meteorites, a characteristic that so far was not clearly elucidated in terms of more fundamental physical
mechanisms. The theory in which we have based on is a
Chern‑Simons-type, which supposes the existence of a
background constant four-vector field that breaks both
Lorentz and parity symmetries, and due to this latter feature, induces a PVED in those molecules, turning an
enantiomeric species energetically more favorable than
the other one. In this work, we have considered a reference frame in which there only is the temporal component of the background field.

We have found in a terrestrial scenario, by considering the action of an modified effective magnetic field on
amino acids in rest on the Earth surface, a maximal
PVED between molecular enantiomers of $10^{-12}$ eV,
implying a difference between the levogyrous and dextrogyrous quantities of the order of $10^{14}$ molecules/mol.

On the other hand, by considering an astronomical
(Jovian) scenario, the referred PVED grows to $10^{-9}$ eV, which corresponds to $10^{-7}$ percent of L-enantiomer
excess, a very low value when compared with the findings in Murchison and Murray meteorites \cite{Cronin}, which may
obviously mean that the mechanism proposed here is not
the only one in action, though it can be the seed for
others.

In Table 1, we compare the calculated PVED above with those of models based on other fundamental interactions and found in literature

We notice that the Carroll‑Field‑Jackiw electrodynamics presents a quite favorable PVED in the scenarios
investigated here, staying within the range of PVED's
predicted in other models involving fundamental interactions. However, if we consider a less stringent value for
$K_0$ ( = $10^{-21}$ GeV or $10^{-6}$ m$^{-1}$) obtained from a terrestrial experiment based on measurements of the Schumann
resonance \cite{Kostelecky}, the PVED values grow to $10^{-4}$
eV (Jovian
scenario) and $10^{-7}$
eV (terrestrial scenario), turning the
former the most favorable bound among the ones originated from fundamental interactions. On the other hand,
if we take into account the astrophysical-based experiments, where $K_0$ = $10^{-42}$ GeV (or $10^{-27}$m$^{-1}$), \cite{Kostelecky} the PVED
are $10^{-25}$ eV and $10^{-28}$ eV, respectively, which we consider
quite hard of observing.
\section*{ACKNOWLEDGEMENTS}

CRM gives thanks to CNPq and FUNCAP for their partial
support, the latter under the grant PRONEM PNE-0112-00085.01.00/16. We also thank Professor Bruno T. O. Abagaro by the useful discussions.

\bibliographystyle{unsrt}  


}

\end{document}